\documentstyle[prl,aps]{revtex}
\begin{document}
\title{Revisiting renormalization schemes in a differential
equation approach}
\author{Ji-Feng Yang}
\address{Department of Physics, East China Normal University,
Shanghai 200062, P R China}
\maketitle
\begin{abstract}
We reconsider the choice of renormalization schemes in a
differential-equation approach to aid the discussion of the
renormalization of the unstable particles and the CKM matrix in
the Standard Model. Certain mass dependent schemes do not satisfy
these natural differential equations modulo trace anomaly. By the
way, the Callan-Symanzik equations were employed to show that both
mass dependent and independent schemes realize fermion decoupling
in the same way.
\end{abstract}
\vspace{0.6cm}
Theoretically, different renormalization schemes are
perfectly equivalent provided the perturbation is completed. The
perturbative scheme dependence is due to the truncation of the
perturbation series\cite{scheme}. Recently, the issue of
renormalization scheme has been intensively investigated in
Standard Model (SM) and its extensions with respect to the
unstable sector (like $W^{\pm}, Z^0$ and Higgs
particles)\cite{polem,Niel} and that of the
Cabibbo-Kobayashi-Maskawa (CKM) matrix\cite{CKM,Gambi}, where
certain mass dependent (like MOM and on-shell) schemes become
problematic in these contexts: the on-shell schemes (mass
dependent) would bring in severe gauge dependence and IR
singularities or ambiguities. The lesson that could be drawn from
these progresses is that the freedom in choosing renormalization
prescriptions deserves closer investigation in applications. In
general, a gauge invariant definition is preferred while gauge
dependent ones should be avoided at all\cite{Niel,Kronfeld}.

In this brief report, we wish to draw attention to other aspects
of the same issue. We start with the standard point of view that
all the known QFT's are effective theories for the low energy
sectors in a completely well-defined quantum theory containing
'correct' high energy details\cite{peskin}. Then we can derive
from this postulate a set of natural differential equations for
the loop amplitudes due to the well-known fact that
differentiating the loop amplitudes with respect to external
parameters like momenta and mass(es) reduces the divergence
degree\footnote{We note that differentiating a Feynman amplitude
with respect to external parameters amounts to physically probing
the quantum process defined by this amplitude, as such operation
($\partial_{p_{\mu}}$ and or $\partial_{ M}$) automatically leads
to the insertion of elementary vertices. Therefore, in this
approach a superficially ill-defined amplitude is
\emph{determined} through its response to physical probes or
'measurements' plus physical boundary conditions. In conventional
renormalization programs, such kind of procedure is effected
through confronting the intermediately renormalized amplitudes
with experiments to obtain physical
parametrization\cite{sterman}.}: the divergent loop amplitudes are
therefore obtained as solutions to these well-defined differential
equations, which are generally finite nonlocal terms plus certain
a polynomial in terms of external momenta, masses and couplings
with arbitrary coefficients to be fixed through physical 'boundary
conditions'\cite{diffeq} as we usually do in quantum mechanics and
electrodynamics. We have demonstrated this simple approach in a
recent study of dynamical symmetry breaking in the two loop
effective potential of massless$\lambda\phi^4$\cite{PRD}.
Obviously, for convergent diagrams such differential equations are
naturally satisfied.

The key point is that, the coefficients in the polynomial, to be
fixed through physical 'boundary conditions', should be
independent of the differentiation parameters like masses. If one
affords some intermediate unphysical definition to these
constants, one could arrive at any renormalization prescription,
like MOM, on-shell and $\overline{MS}$. However, some prescription
might be inconsistent: the original differential equations might
be violated by such prescriptions. Since such differential
equations are well defined, the physical amplitudes must satisfy
these equations and any renormalization prescriptions that solve
these equations could be used. However, it is unclear how those
prescriptions that fail to satisfy such equations could finally
lead to physical amplitudes that satisfy the same equations, as a
practical complete summation of the perturbations series is
impossible.

Now let us demonstrate the differential equation solution of a
loop amplitude with a simple vertex function at the lowest order
in QED, say, the 1-loop photon vacuum polarization tensor\FL
\begin{eqnarray}
\Pi ^{\mu \nu }(p,-p,m)\equiv -ie^2\int d^4 k \ tr\{\gamma ^\mu
\frac 1{ p\!\!\!/+k\!\!\!/-m}\gamma ^\nu \frac 1{k\!\!\!/-m}\}
\end{eqnarray}
in QED for simplicity. Here $m$ refers to the mass of any fermion
with electric charge. We are mainly concerned with the mass
dependence and the differential equation is explicitly given in
terms of mass. In the more interesting case of symmetry breaking
masses, similar analysis should also be feasible and will be
studied in the future.

This amplitude satisfies the following well-defined inhomogeneous
differential equation in any gauge invariant regularization ($GI$)
or in the complete underlying theory, \FL
\begin{eqnarray}
\label{eqa} \partial _m\Pi ^{\mu \nu }(p,-p,m) =-ie^2\int\left(
d^4k\right) _{GI}tr\{\gamma ^\mu (\frac
1{p\!\!\!/+k\!\!\!/-m})^2\gamma ^\nu \frac 1{ k\!\!\!/-m}+\gamma
^\mu \frac 1{p\!\!\!/+k\!\!\!/-m}\gamma ^\nu (\frac 1{
k\!\!\!/-m})^2\}.
\end{eqnarray}Moreover it satisfies the following well defined equation
without regularization at all \FL
\begin{eqnarray}
(\partial _m)^3\Pi ^{\mu \nu}(p,-p,m)=-ie^2\sum_{l=0}^3C_l^3 \int
d^4k\ tr\{\gamma ^\mu (\frac 1{p\!\!\!/+k\!\!\!/-m})^{l+1}\gamma
^\nu (\frac 1{k\!\!\!/-m})^{4-l}\}
\end{eqnarray}
where $C_l^3$ is the combinatorial factor arising from the
differentiation operation.

It suffices to demonstrate our points with Eq.~(\ref{eqa}).
Factorizing out the gauge invariant projector $(g^{\mu \nu
}p^2-p^\mu p^\nu)$ we arrive at the following natural equation\FL
\begin{equation}
\label{eqc}
\partial _m\Pi (p^2,m)=\frac{e^2}{\pi ^2}\int_0^1dx\frac{x(1-x)m}{
m^2-x(1-x)p^2}
\end{equation}
and the solution to this equation reads \FL
\begin{eqnarray}
\Pi ( p^2,m;C)=\frac{e^2}{2\pi ^2} \int_0^1dxx(1-x)\ln
\frac{m^2-x(1-x)p^2}C
\end{eqnarray}
with $C$ being the arbitrary integration constant to be fixed
somehow. Then we have \FL
\begin{eqnarray}
\Pi ^{\mu \nu }(p,-p,m;C)=\frac{e^2}{2\pi^2}(g^{\mu \nu }p^2-p^\mu
p^\nu )\int_0^1dx\ (x-x^2)\ln \frac{m^2-(x-x^2)p^2}C.
\end{eqnarray}Here the main conclusion we
can draw about $C$ is that it must be gauge invariant and mass
independent.

Now we observe that not all renormalization prescriptions preserve
Eq.~(\ref{eqc}). In all mass independent prescriptions, $C$ is
fixed so that the renormalized amplitude still satisfies
Eq.~(\ref{eqc}). However, in the MOM schemes, $C$ is defined in
the following way, \FL
\begin{eqnarray}
\Pi ( p^2,m;C_{MOM}) | _{p^2=-\mu ^2}=0 \Longrightarrow
C_{MOM}=m^2+x(1-x)\mu ^2,
\end{eqnarray}
which leads to the violation of Eq.~(\ref{eqc}),\FL
\begin{eqnarray}
\label{eqd}
\partial _m\Pi ( p^2,m;C_{MOM}) =\frac{e^2}{\pi ^2}
\int_0^1dx \frac{x(1-x)m}{m^2-x(1-x)p^2}+\delta ( \mu^2,m) ,
\end{eqnarray}
with $\delta ( \mu ^2,m) \equiv -\frac{e^2}{\pi ^2} \int_0^1dx
\frac{x(1-x)m}{m^2+x(1-x)\mu ^2}$. This is the main point of our
concern, as such violation might lead to problems in practical
applications, especially in the complicated case like SM and its
extensions. The spirit followed here is the same as is followed in
\cite{Niel,Gambi,Kronfeld} where physical or novel requirement
like gauge invariance are guidelines. In these literature the pole
mass renormalization for unstable particles is therefore
advocated.

There is a subtle point concerned with the pole mass
renormalization for the massless photon, which is just the
on-shell scheme, as $ \Pi ( p^2,m;C)|_{p^2=0}=0\rightarrow C=m^2.
$ Thus it seems that the pole mass definition (or the on-shell
definition) for photons also violates Eq.~(\ref{eqc}). To this end
we note that in this case the physical scale or trace
anomaly\cite{trace} is defined in the following way \FL
\begin{eqnarray}
\{p^{2}\partial_{p^{2}}+m^2\partial_{m^2}-\gamma_A\}[1+\Pi (
p^2,m;C)]=0,\Longrightarrow
\beta(\alpha)=\gamma_A=-2C\partial_C\Pi (
p^2,m;C)|_{C=m^2}=\frac{e^2}{6\pi ^2}=\frac{2\alpha}{3\pi}.
\end{eqnarray}That is, the mass also play the role of the
running scale and the extra term in Eq.~(\ref{eqd}) is now just
the \emph{gauge invariant} trace anomaly with physical
significance--the quantum mechanical violation of naive scaling
law. Then we see that the differential equations could be modified
by physical trace anomaly. From now on we consider the violation
of the differential equations \emph{modulo} trace anomaly.

In the more complicated gauge theories such as SM, due to the
absorptive (threshold effect) part of the self-energy of the
unstable particles, mass dependent schemes could lead to gauge
dependence and inconsistencey\cite{CKM}, i.e., in such theories,
the mass dependent schemes were contaminated with severe gauge
dependence. We expect that in such theories the novel differential
equations proposed here would also be severely violated modulo
gauge invariant trace anomalies. Thus from the differential
equation point of view, combining with the recent
studies\cite{polem,Niel,CKM,Gambi,Kronfeld} severe gauge
dependence and other defects, the mass dependent schemes should be
avoided in computing SM radiative corrections. Theoretically we
could not perform a complete sum of the perturbation series to
test if such defects automatically disappear at all and
practically we have to employ 'good' schemes that are free from
the above mentioned defects before going to experimental data.

In addition, we should note that the pole mass renormalization for
heavy quarks does satisfy the differential equations in terms of
lagrangian masses modulo trace anomalies, as it is gauge invariant
and IR finite\cite{Kronfeld,Tarrach} according to the definition
$m_{eff}(p^2,{\bar m})|_{p^2=M_{pol}^2}=M_{pol}$\cite{Tarrach}. To
see this point we simply note that $m_{eff}$ is a physical 'form
factor' that is scheme independent\cite{Coq}, thus its dependence
upon Lagrangian mass is the same as that in the mass independent
schemes. This argument also applies to the case of massive bosons,
see \cite{polem}.

We can understand the above issues from the underlying theory
point of view. Suppose we could compute everything from the
complete underlying theory, including the 'low energy' amplitudes.
Then in the 'low energy' limits, the 'low energy' amplitudes must
be definite functions (including definite pieces of local terms
that arise from the this limit) in terms of the 'low energy'
physical parameters. There must also be a reference scale for
specifying the relevant 'low energy' processes. Then the one loop
photon polarization should take the following form\FL
\begin{eqnarray}
\Pi^{\mu \nu}(p^2,m_{u.t.};\{\sigma\}) |_{\{\sigma\}\Rightarrow 0
}= \frac{e^2_{u.t.}}{\pi ^2}(g^{\mu \nu }p^2-p^\mu p^\nu
)\int_0^1dx x(1-x)\ln \frac{m_{u.t.}^2-x(1-x)p^2}{\mu^2_{u.t.}},
\end{eqnarray}with the subscript 'u.t.' referring to the
underlying theory and $\{\sigma\}$ denoting the fundamental
constants of the underlying theory. It is natural to identify the
'low energy' parameters like $m_{u.t.}$ with the Lagrangian
parameters of QFT (bare but finite). The scale $\mu_{u.t.}$, being
process specific and physical and hence renormalization group (RG)
and renormalization scheme (RS) invariant, is not the running
scale though it appears in the same place. Then any
renormalization prescriptions that are equivalent to such physical
parametrization is acceptable. However, the difficulty is, we do
not know how to identify such process dependent scale
theoretically before the complete underlying theory is known,
therefore we must resort to other means like experimental data and
some reasonable or physical requirements. Before the complete
theory is found,we could not exclude the possibility that there
might be schemes that are physically inequivalent to the complete
theory definitions at all.

Moreover, from the differential equation point of view, the scheme
invariance of certain quantities in QFT originally established
within the mass independent schemes\cite{scheme} could be promoted
to a full 'invariance' for all the schemes that satisfy the
differential equations in terms of all physical parameters.
Otherwise, if the mass dependent schemes also qualified for
applications, the significance of such 'invariance' established
within the mass independent schemes is diminished in the cases
where masses could not be neglected. In a sense we put forward an
argument for advocating mass independent schemes, and this point
of view could draw supports from the recent literature on
renormalization schemes, though the relation between our approach
here and those in the literature is unclear at this moment.

Before closing the presentation we would like to discuss the issue
of heavy particles decoupling where the MOM and other mass
dependent schemes are thought to be advantageous over the mass
independent schemes due to its good decoupling behavior. We first
note that this advantage is only in the context of RGE. Then we
also remind that, as we will demonstrate below, in the context of
Callan-Symanzik equations\cite{CS} (CSE), which describe the full
scale behavior, the decoupling of heavy fields\cite{AC} is
achieved in the same way in all schemes.

Again we illustrate it with a simple model, QED with a massive
fermion in addition to $n_l$ massless fermions. In a mass
independent scheme the Callan-Symanzik equation reads, \FL
\begin{eqnarray}
\{\lambda \partial_\lambda -\beta \alpha \partial _\alpha +\gamma
_\Gamma -D_\Gamma \}\Gamma ((\lambda p),m,\alpha ,\mu ) = -i\Gamma
^\Theta ((\lambda p),m,\alpha ,\mu )
\end{eqnarray}
where $\Theta \equiv {[1+\gamma_m]}m{\bar{\psi}}\psi $, $\beta $,
$\gamma _\Gamma $ and $\gamma _m$ are mass independent functions
of the renormalized coupling $\alpha $ and all quantities are
renormalized ones. At lowest order, $\beta =\frac{2\alpha }{3\pi
}{[n_l+1]}$.

When the mass goes to infinity, $\beta,\gamma$ on the left hand
side of CSE does not change at all, but in the meantime, the
inhomogeneous term of CSE does not vanish,\FL
\begin{eqnarray}
&&\label{eqdec1} -i\Gamma^\Theta ((p),m,\alpha ,\mu
)|_{m\rightarrow \infty } =(\Delta \beta \alpha \partial _\alpha
-\Delta \gamma _\Gamma
)\Gamma_0 ((p),\alpha ,\mu ),\\
\label{eqdec} &&\Longrightarrow \{\lambda
\partial_\lambda -\beta _{0}\alpha \partial _\alpha +\gamma
_{\Gamma ;0}-D_{\Gamma} \}\Gamma_0((\lambda p),\alpha ,\mu )=0.
\end{eqnarray}
Here $\beta_{0}\equiv \beta +\Delta \beta,\ \gamma _{\Gamma
;0}\equiv \gamma _\Gamma +\Delta \gamma _\Gamma $ with the delta
contributions coming from the mass insertion part in the infinite
mass limit which will cancel the heavy field's contribution to
$\beta $ and $\gamma $. The subscript 0 means that the heavy
particle is removed. The generalization to other theories with
boson masses is an easy exercise. From Eq.~(\ref{eqdec}) we see
that the decoupling of heavy particles is realized in a natural
way in the contexts of Callan-Symanzik equations.

To verify the above deduction it is enough to demonstrate
Eq.~(\ref{eqdec1}) at the lowest order which is closely related to
the observation that heavy particle limit provides a convenient
algorithm for calculating trace anomalies\cite{traceyang}: \FL
\begin{eqnarray}
\label{eqM} -i\langle m{\bar{\psi}}\psi J^\mu J^\nu \rangle
|_{m\rightarrow \infty }= \frac{2\alpha }{3\pi }(p^2g^{\mu \nu
}-p^\mu p^\nu)\Rightarrow& m(1+\gamma _m){\bar{\psi}}\psi \Vert
_{m\rightarrow \infty }=\frac 14\Delta \beta F^{\mu \nu }F_{\mu
\nu },\ \
\end{eqnarray}
with $\ J^\mu \equiv -ie{\bar{\psi}}\gamma ^\mu \psi $ and $\Delta
\beta \equiv -\frac{2\alpha }{3\pi }=\Delta \gamma _A$. When
translated into Callan-Symanzik equations Eq.~(\ref{eqM}) is just
Eq.~(\ref{eqdec1}). The cancellation of the heavy particle
contributions is obvious since $\beta +\Delta \beta =\frac{
2\alpha }{3\pi }{[n_l+1]}-\frac{2\alpha }{3\pi }=\frac{2\alpha
}{3\pi }n_l$.

While in the MOM like schemes the Callan-Symanzik equation reads,
\FL
\begin{eqnarray}
\{\lambda \partial _\lambda -\beta_{MOM}\alpha \partial _\alpha
+\gamma _{\Gamma;MOM}-D_\Gamma \}\Gamma_{MOM}((\lambda p),m,\alpha
,\mu )=-i\Gamma_{MOM}^\Theta ((\lambda p),m,\alpha ,\mu)
\end{eqnarray}
with the beta function etc. being defined as $\ \beta_{MOM}\equiv
[\mu
\partial _\mu +m(1+\gamma_{m;MOM})\partial _m]\alpha ,\cdots ,$ in
contrast to the RGE definition: $\beta _{MOM}^{RG}\equiv \mu
\partial _\mu \alpha ,\cdots ,$ due to the mass dependence of the
renormalization constants\footnote{In the standard inhomogeneous
form of Callan-Symanzik equation, the mass operators inserted
vertex functions appear in the other side of the equation and the
definitions of $\beta,\gamma$ in the MOM-like mass dependent
schemes must include $m(1+\gamma_{m;MOM})\partial _m$ to account
for the 'full running' of the renormalization constants. But in
the alternative homogeneous form of Callan-Symanzik equation
(CSE), i.e., $ \{\lambda \partial _\lambda -\beta\alpha \partial
_\alpha +m(1+\gamma_m)\partial _m+\gamma _{\Gamma}-D_\Gamma
\}\Gamma((\lambda p),m,\alpha ,\mu )=0$, the definitions of the
$\beta,\gamma$ etc. are the same as in RGE.}. The resulting
$\beta_{MOM}$ also exhibits non-decoupling feature as in the mass
independent schemes. For example, at the lowest order, from the
definition given above, a heavy particle's contribution to $\beta$
at lowest order is mass independent \FL
\begin{eqnarray}
\label{betaMOM} \beta_{MOM;M}=(\mu \partial _\mu
+m(1+\gamma_{m;MOM})\partial _m)
\frac{e^2}{2\pi ^2}%
\int_0^1dxx(1-x)\ln C_{MOM}= \frac{2\alpha }{3\pi }.
\end{eqnarray}
Without any doubt such 'non-decoupling' term is cancelled by the
contribution from the decoupling limit of the inhomogeneous term
in CSE, that is, just like in Eq.~(\ref{eqdec1}), we have \FL
\begin{eqnarray}
-i\Gamma_{MOM}^\Theta ((p),m,\alpha ,\mu )|_{m\rightarrow \infty }
=(\Delta \beta_{MOM}\alpha \partial _\alpha -\Delta
\gamma_{\Gamma;MOM})\Gamma_{MOM;0}((p),\alpha ,\mu),
\end{eqnarray}
with $\Delta \beta_{MOM}=-\frac{2\alpha }{3\pi }$, which exactly
cancels the heavy particle's contribution to the $\beta,\gamma$
etc. Thus we proved that Eq.~(\ref{eqdec1}) is also true in the
MOM schemes at the lowest order. It is well known that the first
loop order beta of RGE in mass independent schemes differs from
that in the MOM schemes\cite{Coq}. While in the context of
Callan-Symanzik equation the beta function is the same in all
schemes at one loop order (C.f. Eq.~(\ref{betaMOM})), which in
turn implies that the same decoupling mechanism works in both mass
independent and mass dependent schemes in the context of
Callan-Symanzik equation. Of course, in the preceding derivation
it is assumed that there is no such trouble as gauge dependence
and/or any form of inconsistency for the mass dependent schemes,
i.e., we consider the theories that are not beset with the
troubles that afflicts the electroweak sector of SM.

In summary, we revisited the issue of renormalization schemes that
is intensely discussed in the literature of Standard Model. The
analysis, basing on a set of natural differential equations, seems
to favor the mass independent schemes against the mass dependent
ones, supporting the point of view of the recent literature
concerning the renormalization of CKM matrix and the unstable
sectors of SM. It is also demonstrated that in the context of
Callan-Symanzik equation all the schemes facilitate the decoupling
of heavy particles in the same way. Further investigation in this
direction seems worthwhile.
\section*{Acknowledgement}
The author is grateful to W. Zhu for useful discussions. This work
is supported by the National Nature Science Foundation of China
under Grant No. 10075020 and No. 10502004.


\begin{references}
\bibitem{scheme} P. M. Stevenson, Phys. Rev. {\bf D23}, 2916 (1981);
G. Grunberg, Phys. Rev. {\bf D29}, 2315 (1984); S. J. Brodsky, G.
P. Lepage and P. B. Mackenzie, Phys. Rev. {\bf D28}, 228(1983); D.
T. Barclay, C. J Maxwell and M. T. Reader, Phys. Rev. {\bf D49},
3480 (1994).
\bibitem{polem} See, e.g., S. Willenbrock and G. Valencia, Phys.
Lett. {\bf B259}, 373(1991); R. G. Stuart, Phys. Lett. {\bf B262},
113 (1991); A. Sirlin, Phys. Rev. Lett. {\bf 67}, 2127 (1991),
Phys. Lett. {\bf B267}, 240(1991); T. Bhattacharya and S.
Willenbrock, Phys. Rev. {\bf D47}, 4022 (1993); H. Veltman, Z.
Phys. {\bf C62}, 35 (1994); M. Passera and A. Sirlin, Phys. Rev.
Lett. {\bf 77}, 4146 (1996), Phys. Rev. {\bf D58}, 113010 (1998);
B. A. Kniehl and A. Sirlin, Phys. Rev. Lett. {\bf 81}, 1373( 1998)
and references therein.

\bibitem{Niel} P. Gambino and P. A. Grassi, Phys. Rev.
\textbf{D62}, 076002 (2000).
\bibitem{CKM} See, e.g., A. Denner and T.
Sack, Nucl. Phys. \textbf{B347}, 203 (1990);  B. A. Kniehl, F.
Madricardo, M. Steinhauser, Phys. Rev. \textbf{D62}, 073010
(2000); A. Barroso, L. Brucher and R. Santos, Phys. Rev.
\textbf{D62}, 096003 (2000); Y. Yamada, Phys. Rev. \textbf{D64},
036008 (2001); K. P. Diener and B. A. Kniehl, Nucl. Phys.
\textbf{B617} , 291 (2001) and references therein.
\bibitem{Gambi} P. Gambino, P. A. Grassi and F. Madricardo, Phys.
Lett. \textbf{B454}, 98 (1999);D. Espriu, J. Manzano and P.
Talavera, Phys. Rev. \textbf{D66}, 076002 (2002) and references
therein.
\bibitem{Kronfeld} A. S. Kronfeld, Phys. Rev. {\bf D58}, 051501
(1998).
\bibitem{peskin}See, e.g., M. E. Peskin and D. V.
Schroeder, {\em An Introduction to Quantum Field Theory},
(Addison-Wesley, 1995), Chapter 8 and S. Weinberg, {\em The
Quantum Theory of Fields}, Volume I, (Cambridge University Press,
Cambridge, 1995), Chapter 12.
\bibitem{diffeq} Ji-Feng Yang, arXiv: hep-th/9708104; invited talk,
in {\it Proceedings of the 11th International Conference:
'PQFT98'}, Ed. by B. M. Barbashov, {\em et al}, (Publishing
Department of JINR, Dubna, 1999), 202-206[arXiv: hep-th/9901138];
arXiv: hep-th/9904055; arXiv: hep-ph/0212208.
\bibitem{PRD} Ji-Feng Yang and Jian-Hong Ruan, Phys. Rev.
\textbf{D65}, 125009 (2002).
\bibitem{sterman} G. Sterman, {\em An Introduction to Quantum Field
Theory}, (Cambridge University Press, 1993), Chapter 10.
\bibitem{trace} S. L. Adler, J. C. Collins and A. Duncan, Phys. Rev.
\textbf{D15}, 1712 (1977); J. C. Collins, A. Duncan and S. D.
Joglekar, Phys. Rev. \textbf{D16}, 438 (1977).

\bibitem{Tarrach} R. Tarrach, Nucl. Phys. {\bf B183}, 384 (1981);
N. Gray, {\it et al}, Z. Phys. {\bf C48}, 673 (1990).

\bibitem{Coq} See, e.g., R. Coquereaux, Ann. Phys. {\bf 125}, 401( 1980).
\bibitem{CS} C. G. Callan, Jr., Phys. Rev. {\bf D2},1541 (1970); K.
Symanzik, Comm. Math. Phys. {\bf 18}, 227 (1970).

\bibitem{AC} T. Appelquist and J. Carazzone, Phys. Rev. {\bf D11}, 2856
(1975).
\bibitem{traceyang} Ji-Feng Yang, Ph D Thesis, Fudan University, unpublished,
(1994); G.-j. Ni and Ji-Feng Yang, Phys. Lett. {\bf B393}, 79
(1997).
\end{references}
\end{document}